\documentclass[a4paper,11pt]{article}
\usepackage{pos}
\usepackage{siunitx}
\usepackage{caption}
\usepackage{subcaption}

\title{Status of the muEDM experiment at PSI}

\author*[a]{Francesco Renga}
\onbehalf{on behalf of the muEDM collaboration}

\affiliation[a]{INFN Sez. di Roma,\\
  P.le A. Moro 2, 00185 Roma, Italy}

\emailAdd{francesco.renga@roma1.infn.it}

\abstract{Explaining the matter-antimatter asymmetry in the Universe requires new sources of CP violation beyond the predictions of the Standard Model (SM). Electric dipole moments (EDMs) of particles, being zero if CP is exactly conserved and extremely small in the SM, are a very clean and sensitive probe for new physics. We will present the status of the muEDM experiment, a search for a muon EDM at PSI (CH) pioneering the frozen spin technique. Muons will be stored in a solenoid, with a radial electric field tuned to eliminate the spin precession generated by the magnetic moment. Measuring a residual, longitudinal precession would indicate a non-zero EDM. The first phase of the experiment will demonstrate, by 2026, the feasibility and unique potential of the technique, while reaching a sensitivity competitive with the parasitic measurements performed in the muon $g-2$ experiments. The ultimate goal of the muEDM experiment is to improve this sensitivity by a factor of 100 by the early 2030s.}

\FullConference{42nd International Conference on High Energy Physics (ICHEP2024)\\
18-24 July 2024\\
Prague, Czech Republic\\}

\begin{document}
\maketitle

\section{Introduction}

A fundamental particle having a non-null electric dipole moment (EDM) $d$ would interact with electric fields generating terms of the form $H_{EDM} = - d(\vec S \cdot \vec E)$. Under time reversal, the spin $\vec S$ changes of sign, while the electric field does not, resulting in a T-symmetry violation that, assuming CPT invariance, implies a violation of CP.

In the Standard Model (SM), the CP violation effects in the weak interactions of quarks produce only small EDMs for barions, and even smaller EDMs for leptons through many-loop interactions. For example, the EDM is expected to be around $10^{-32}~e \cdot \mathrm{cm}$ for the neutron~\cite{Dar:2000tn}, $10^{-40}~e \cdot \mathrm{cm}$ for the electron and $10^{-38}~e \cdot \mathrm{cm}$ for the muon~\cite{Yamaguchi:2020eub}. However, CP violation sources larger than in the SM are necessary to explain the matter-antimatter asymmetry observed in the Universe. It motivates the search for an enhancement of the EDMs of fundamental particles, as predicted by several new physics models~\cite{Raidal:2008jk}.

From an experimental point of view, the EDMs of leptons predicted by the SM are too low to be observed in the near future, the current experimental limits on the electron and muon EDMs being, at 90\% confidence level~\cite{Roussy:2022cmp,Muong-2:2008ebm}:
\begin{eqnarray}
d_e &<& 4.1 \times 10^{-30}~e \cdot \mathrm{cm} \; , \\ 
d_\mu &<& 1.9 \times 10^{-19}~e \cdot \mathrm{cm} \; .
\end{eqnarray}
While the electron limit comes from a dedicated experiment on electrons in intramolecular electric fields, the muon limit was produced by the BNL Muon $g-2$ collaboration as a byproduct of their muon $g-2$ measurement campaign. The FNAL $g-2$ experiment is expected to improve this limit by a factor 20~\cite{Foster:2023kzz}.

The goal of the muEDM collaboration is to perform a dedicated experiment for the search of a non-null muon EDM, adopting a frozen-spin technique~\cite{Adelmann:2010zz}, down to a few $10^{-23}~e \cdot \mathrm{cm}$~\cite{Adelmann:2021udj}. It is worth mentioning that, under the assumption of minimal flavor violation (MFV) in physics beyond the SM, the existing limit on the electron EDM implies a constraint on the muon EDM that is orders of magnitude below this experimental reach. However, the existing tensions in flavor physics disfavor the MFV scenario, leaving room for a muon EDM as large as the experimental sensitivity (see~\cite{Adelmann:2021udj, Crivellin:2018qmi} for a more detailed discussion, and references therein).

\section{The frozen-spin technique}

Measurements of the muon dipole moment rely on the analysis of the spin dynamics of muons moving with $\vec \beta =\vec v/c$ in electric ($\vec E$) and magnetic ($\vec B$) fields, which is described by the well-known Thomas-BMT equation:
\begin{align}
	\vec{\Omega}=&\frac{q}{m}\left[a\vec{B}-\frac{a\gamma}{\left(\gamma+1\right)}\left(\vec{\beta}\cdot\vec{B}\right)\vec{\beta}-\left(a+\frac{1}{1-\gamma^2}\right)
	\frac{\vec{\beta}\times\vec{E}}{c}\right] \nonumber \\
	&+\frac{\eta q}{2m}\left[\vec{\beta}\times\vec{B}+\frac{\vec{E}}{c}-\frac{\gamma c}{(\gamma+1)}\left(\vec{\beta}\cdot\vec{E}\right)\vec{\beta}\right] \; ,
\label{eq:omegaMuWithEDM}
\end{align}
where $\vec{\Omega}$ is the difference between the spin precession and cyclotron frequencies, $q$ and $m$ are the muon charge and mass, and $a =(g-2)/2$ and $d = \eta e \hbar / (4mc)$ are the anomalous magnetic moment and the EDM of the muon. 

Given the polarization of muons produced in pion decays and the correlation between the muon spin and the kinematics of its decay, the spin precession can be analyzed by looking at the energy and momentum distributions of the decay electrons versus time.

For the measurement of the anomalous magnetic moment, experiments like the past BNL and present FNAL ones use muons with the "magic" momentum $p_\mathrm{magic} = m/\sqrt{a} = 3.09~\mathrm{GeV}/c$, and neglect the EDM contribution, resulting in the simplified equation:
\begin{equation}
\vec\Omega = \frac{q}{m} a \vec B \; ,
\end{equation}
while, if the EDM contribution is not neglected, the precession plane is tilted, the precession frequency is slightly increased and the experiment gets some sensitivity to the EDM.

For a dedicated measurement of the muon EDM, we propose adopting a combination of electric field, magnetic field and muon momentum such that the magnetic precession term in Eq.~\ref{eq:omegaMuWithEDM} is completely canceled and the spin is "frozen" in the direction of the muon momentum. The condition to be satisfied is:
\begin{equation}
		a\vec{B} = \left(a-\frac{1}{\gamma^2-1}\right)\frac{\vec{\beta}\times\vec{E}}{c} \; ,
\label{eq:FrozenSpinCondition}
\end{equation}
and, for $\vec \beta \cdot \vec B = \vec \beta \cdot \vec E = 0$ and $\vec B = \vec E = 0$, the required electric field is $E_f \sim aBc\beta\gamma^2$.
This values turns out to be much smaller than the $c\vec \beta \times \vec B$, and the precession formula simplifies to:
\begin{equation}
\vec\Omega = \frac{\eta q}{2m} \, ( \beta \times \vec B ) \; .
\end{equation}
It is interesting to notice that the residual precession is not dominated by the interaction of the spin with the external electric  $E_f$, but with the electric field $c\vec \beta \times \vec B$ that is observed by the muon in its rest frame, due to relativistic effects.

In this configuration, a non-null EDM generates a precession of the spin in the direction orthogonal to the muon orbit plane, which can be observed by looking at the rate and energy spectrum of electrons emitted toward the two sides of the plane (up and down). In the simplest approach, an asymmetry is measured between the rates of upward and downward electrons above a certain energy threshold, oscillating with time. For an EDM around the current sensitivities, the oscillation period is much larger than the muon decay lifetime, and only the initial buildup of the asymmetry, approximately linear with time, can be observed:
\begin{equation}
A(t) = \frac{N_\mathrm{up} - N_\mathrm{down}}{N_\mathrm{up} + N_\mathrm{down}} \sim \frac{2 P_0 E_f \alpha |d_\mu|}{a \hbar \gamma^2} \, t \; ,
\end{equation}
where $P_0$ is the initial degree of polarization and $\alpha < 1$ is a factor, depending on the energy threshold, indicating homaw much the polarization translates into a detection asymmetry. The resulting sensitivity on the EDM, considering a measurement over one muon lifetime in the laboratory frame $\gamma \tau_\mu$, is approximately:
\begin{equation}
\sigma(d_\mu) \sim \frac{a \hbar \gamma}{2 P_0 E_f \sqrt{N} \tau_\mu \alpha} \; .
\end{equation}
It can be generalized to other analysis techniques, exploiting in more detail the dependence of the decay kinematics on the spin, with $\alpha$ representing the "spin analysis power" of the specific technique.

\section{Conceptual design of the muEDM experiment}

The high-intensity proton accelerator facility at the Paul Scherrer Institut (PSI, Switzerland) delivers the most intense continuous muon beams in the world,  with energies below a few hundred MeV, which are ideal for performing a muon EDM search with the frozen-spin technique. The 125~MeV/$c$ positive muons of the $\mu E1$ beamline can be stored in a 3~T solenoid magnet, with a radial electric field $E_f = 1.92~$MV/m to ensure the frozen spin condition. Injecting muons from one end of the solenoid and capturing them in a stable orbit at the center will require a magnetically shielded injection channel, a weak focusing field with a minimum at the center of the solenoid, and a magnetic kick synchronized with the arrival of the muon. Considering the beam phase space, state-of-the-art superconductive channels, and a properly designed kicker, simulations predict a 0.3\% capture efficiency. With a maximum beam rate of $1.2 \times 10^8~\mu$/s, a storage rate of $3.6 \times 10^5$~s$^{-1}$ would be obtained.

\begin{figure}[htbp]
\begin{center}
\includegraphics[width=0.45\textwidth]{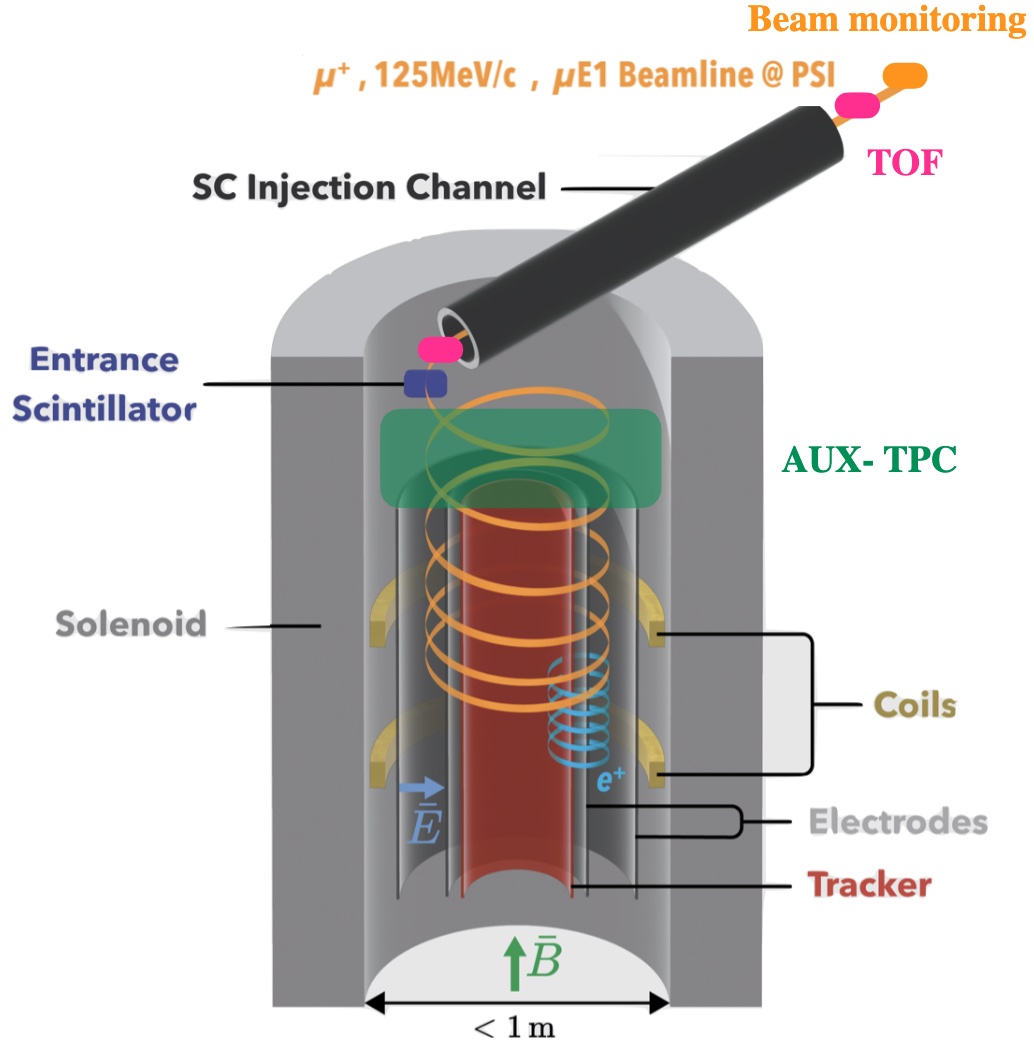}
\caption{A sketch of the muEDM experimental layout.}
\label{fig:layout}
\end{center}
\end{figure}

A conceptual design of the experiment is shown in Fig.~\ref{fig:layout}. Muons enter the solenoid through the superconducting injection channel, which is designed to transmit only muons close to the nominal trajectory for being captured in the center of the solenoid. A detector made of thin scintillators is placed at the exit of the channel and is used to trigger the magnetic kick when a muon arrives. To minimize the dead time, this detector is used in anti-coincidence with another set of scintillators, with such an arrangement that only muons with a very high probability of being captured would pass through without hitting them. At the center of the solenoid, two cylindrical electrodes create the electric field that is necessary to freeze the muon spin. The stored muons will decay into positrons, the trajectory of which will be reconstructed by a set of detectors (silicon pixel detectors and scintillating fibers) surrounding the orbit.

In a one-year run, considering the expected positron acceptance (25\%) and the typical analysis power ($\alpha \sim 0.3$) and initial polarization ($P
_0 \sim 95\%$), a statistical sensitivity of $6 \times 10^{-23}$ is foreseen, two orders of magnitude below the sensitivity of the FNAL $g-2$ experiment.

Systematic uncertainties mainly arise from imperfections of the electric and magnetic field~\cite{Cavoto:2023xtw}. Thanks to the CP-conserving nature of electromagnetic interactions, the most relevant systematics can be canceled by injecting the muons in clockwise and counterclockwise orbits and taking the average of the measured EDMs, if a sufficient symmetry between the two orbits is guaranteed.

\section{The phase-1 experiment}

In order to demonstrate the feasibility of this muEDM search, a phase-1 experiment will be performed, exploiting an existing magnet on the $\pi E1$ beamline with $4 \times 10^6~\mu$/s at 28~MeV/$c$ momentum. The main goal of this experiment is to demonstrate the feasibility of the injection scheme and the frozen-spin condition. However, a storage rate of $2 \times 10^3$~s$^{-1}$ is expected, resulting in an EDM sensitivity of about $3 \times 10^{-21}~e \cdot \mathrm{cm}$, still competitive with the upcoming limit from the FNAL $g-2$ experiment.

\begin{figure}[t]
\centering 
\begin{subfigure}[b]{0.68\textwidth}
	\centering
	\includegraphics[width=\textwidth]{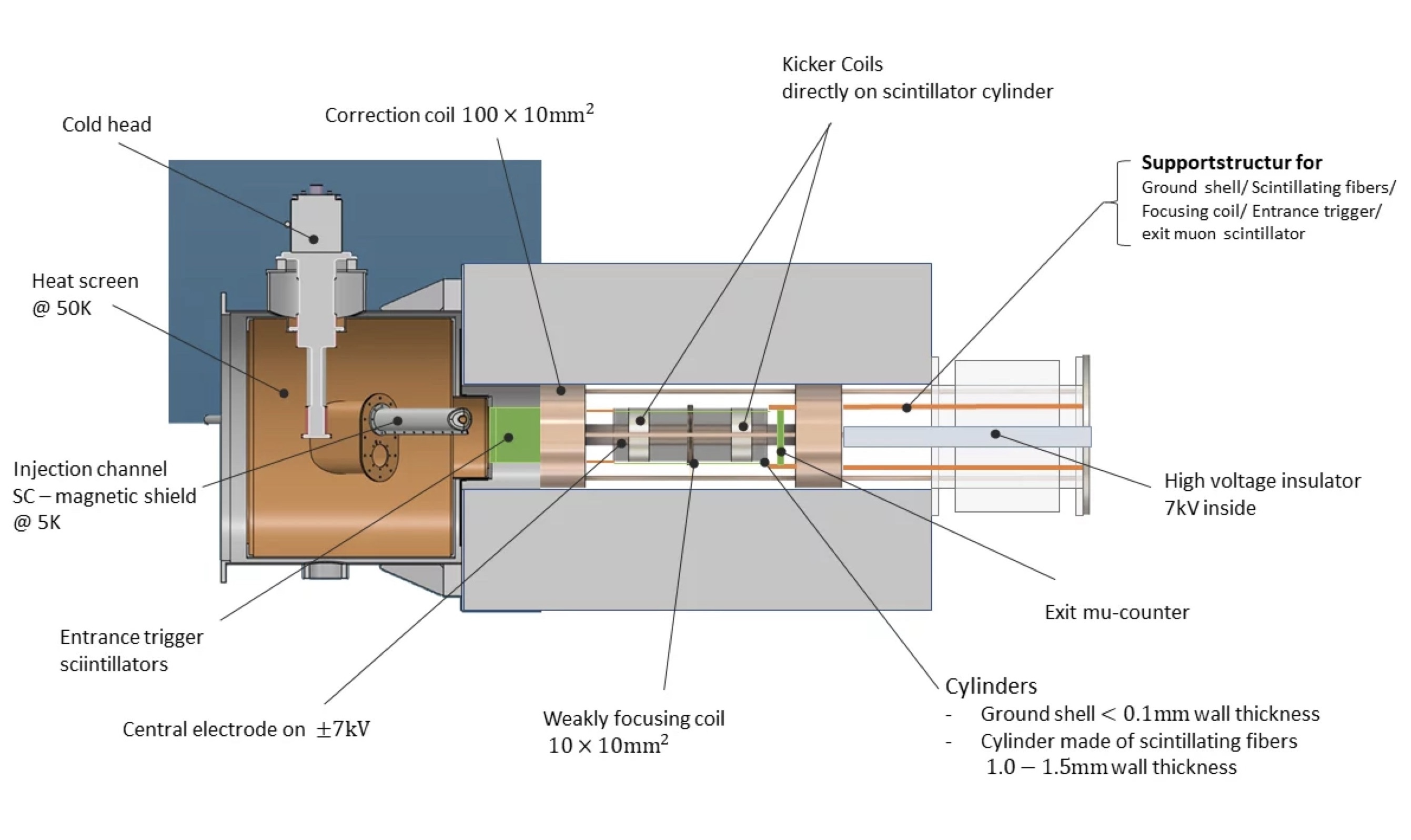}
	\hspace{3cm}
	\caption{}
\end{subfigure}
\begin{subfigure}[b]{0.23\textwidth}
	\centering
	\includegraphics[width=\textwidth]{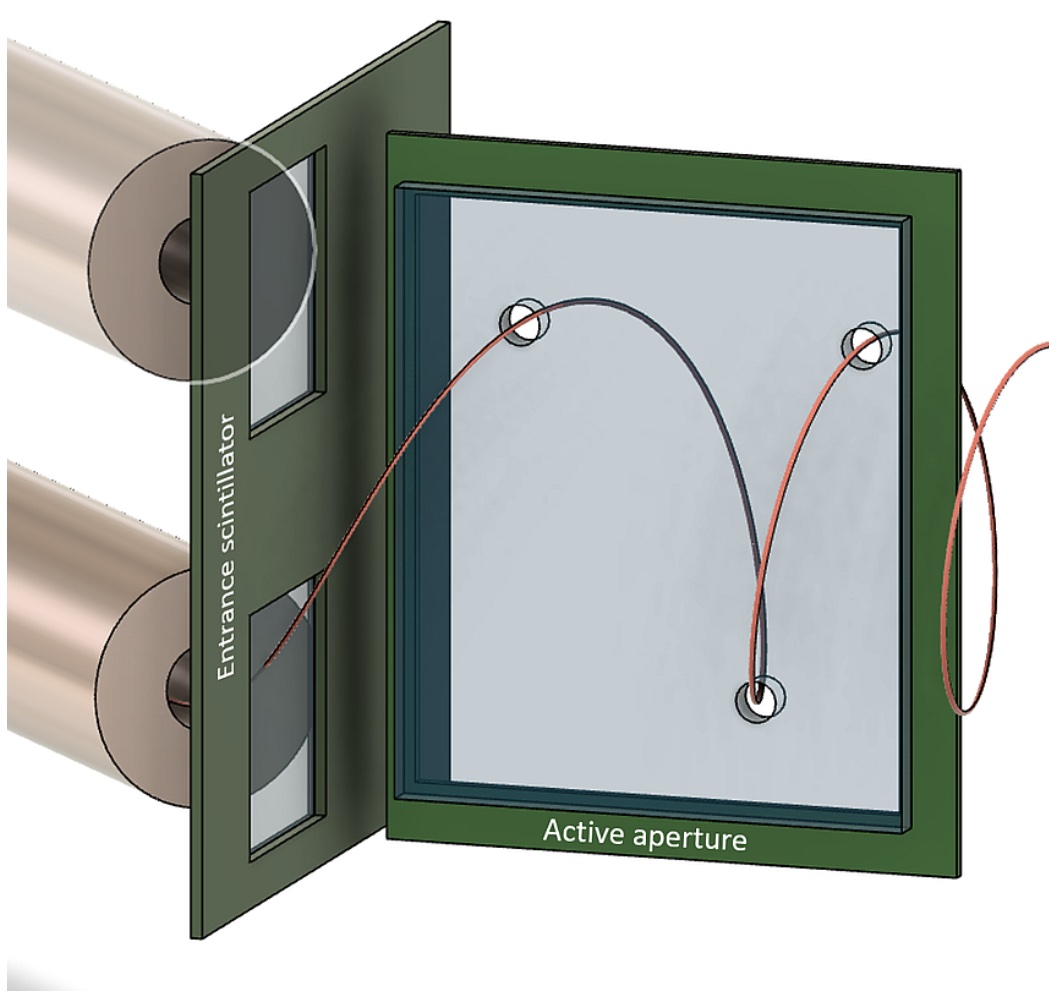}
	\includegraphics[width=\textwidth]{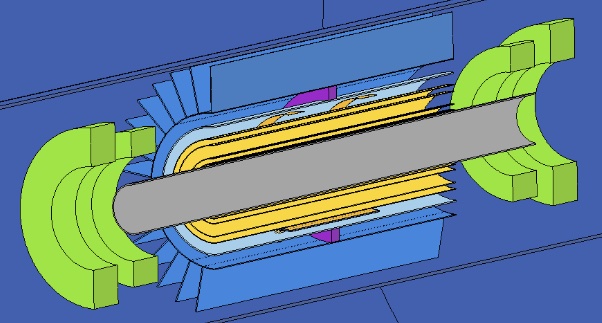}
	\includegraphics[width=\textwidth]{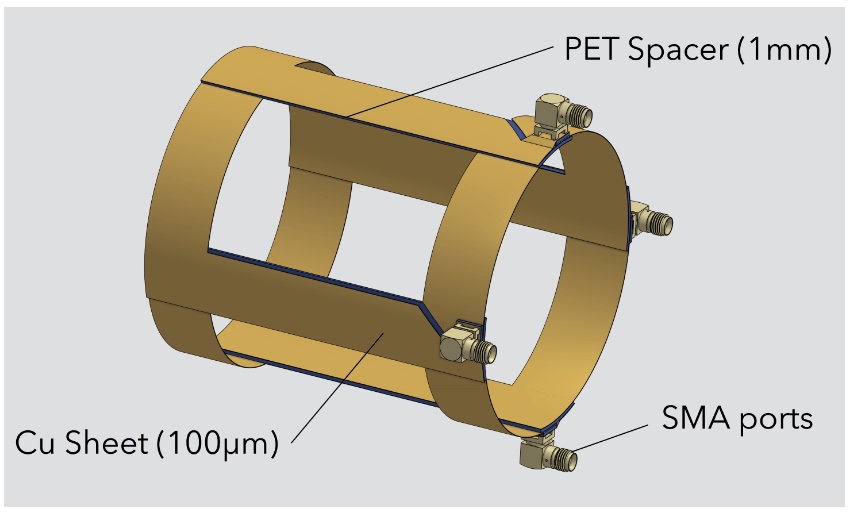}
	\caption{}
\end{subfigure}
\caption{A sketch of the muEDM phase-1 experiment (a), with the current design of the entrance trigger detector (b, top), the positron tracker (b, middle), and the magnetic kicker (b, bottom).}
\label{fig:precursor}
\end{figure}

The experiment will feature a simplified arrangement of detectors. In particular, the positron tracker will be composed only of about 1000 scintillating fibers, arranged to provide 3-dimensional tracking to verify the cancellation of the $g-2$ precession and measure a possible EDM-induced asymmetry. Figure~\ref{fig:precursor} shows a sketch of the experimental setup and the current design of some components of the apparatus: the entrance trigger detector, the positron tracker, the magnetic kicker.

The experiment will also include ancillary detectors for the commissioning of the beam injection system and the control of the systematic uncertainties, namely a time-of-flight detector for the measurement of the average beam momentum and an extremely light gaseous time projection chamber for a detailed characterization of the muon trajectory at the entrance of the solenoid.

\section{Conclusions}

A dedicated experiment to search for a non-null muon EDM is being realized at the Paul Scherrer Institut. The experiment aims to reach a sensitivity of two orders of magnitude better than the current and near-future experiments, which are primarily designed to measure the muon $g-2$.

The final search will be anticipated by a phase-1 experiment, which is under construction and is planned to take data in 2026. The construction of the final, phase-2 experiment will be started soon after, and the data taking is expected to take place in the early 2030s.


\begin{thebibliography}{99}

\bibitem{Dar:2000tn}
S.~Dar,
\textit{The Neutron EDM in the SM: A Review},
[arXiv:hep-ph/0008248 [hep-ph]].

\bibitem{Yamaguchi:2020eub}
Y.~Yamaguchi and N.~Yamanaka,
\textit{Large long-distance contributions to the electric dipole moments of charged leptons in the standard model},
Phys. Rev. Lett. \textbf{125} (2020), 241802.

\bibitem{Raidal:2008jk}
M.~Raidal \textit{et al.},
\textit{Flavour physics of leptons and dipole moments},
Eur. Phys. J. C \textbf{57} (2008), 13-182.

\bibitem{Roussy:2022cmp}
T.~S.~Roussy \textit{et al.},
\textit{An improved bound on the electron\textquoteright{}s electric dipole moment},
Science \textbf{381} (2023) no.6653, adg4084.

\bibitem{Muong-2:2008ebm}
G.~W.~Bennett \textit{et al.} [Muon (g-2)],
\textit{An Improved Limit on the Muon Electric Dipole Moment},
Phys. Rev. D \textbf{80} (2009), 052008.


\bibitem{Foster:2023kzz}
S.~B.~Foster \textit{et al.},
\textit{Muon $g-2$ and EDM at Fermilab},
PoS \textbf{Muon4Future2023} (2024), 016.

\bibitem{Adelmann:2010zz}
A.~Adelmann, K.~Kirch, C.~J.~G.~Onderwater and T.~Schietinger,
\textit{Compact storage ring to search for the muon electric dipole moment},
J. Phys. G \textbf{37} (2010), 085001.

\bibitem{Adelmann:2021udj}
A.~Adelmann \textit{et al.},
\textit{Search for a muon EDM using the frozen-spin technique},
[arXiv:2102.08838 [hep-ex]].

\bibitem{Crivellin:2018qmi}
A.~Crivellin, M.~Hoferichter and P.~Schmidt-Wellenburg,
\textit{Combined explanations of $(g-2)_{\mu,e}$ and implications for a large muon EDM},
Phys. Rev. D \textbf{98} (2018) no.11, 113002.

\bibitem{Cavoto:2023xtw}
G.~Cavoto  \textit{et al.},
\textit{Anomalous spin precession systematic effects in the search for a muon EDM using the frozen-spin technique},
Eur. Phys. J. C \textbf{84} (2024) no.3, 262. 

\end{thebibliography}
\end{document}